\documentclass[apj,iop]{emulateapj}

\usepackage{natbib}
\usepackage{graphicx}
\usepackage{dcolumn}
\usepackage{bm}
\usepackage{amsmath}
\usepackage{amssymb}
\usepackage{hyperref}
\usepackage{xcolor}

\shorttitle{The effect of radial migration on galactic disks}
\shortauthors{Vera-Ciro et al.}

\hypersetup{
    bookmarks=true,              
    unicode=false,               
    pdftoolbar=true,             
    pdfmenubar=true,             
    pdffitwindow=false,          
    pdfstartview={FitH},         
    pdftitle={Rad Migration},    
    pdfauthor={Vera-Ciro et al.},
    pdfnewwindow=true,           
    colorlinks=true,             
    linkcolor=blue,              
    citecolor=blue,              
    filecolor=blue,              
    urlcolor=blue                
  }

\begin{document}

\title{The effect of radial migration on galactic disks}

\author{Carlos Vera-Ciro\altaffilmark{1}, Elena D'Onghia\altaffilmark{1,4},
  Julio Navarro\altaffilmark{2,5}, Mario Abadi\altaffilmark{3}}

\begin{abstract}
  We study the radial migration of stars driven by recurring multi-arm spiral
  features in an exponential disk embedded in a dark matter halo. The spiral
  perturbations redistribute angular momentum within the disk and lead to
  substantial radial displacements of individual stars, in a manner that
  largely preserves the circularity of their orbits and that results, after
  $5$ Gyr ($\sim 40$ full rotations at the disk scalelength), in little radial
  heating and no appreciable changes to the vertical or radial structure of
  the disk. Our results clarify a number of issues related to the spatial
  distribution and kinematics of migrators. In particular, we find that
  migrators are a heavily biased subset of stars with preferentially low
  vertical velocity dispersions. This ``provenance bias'' for migrators is not
  surprising in hindsight, for stars with small vertical excursions spend more
  time near the disk plane and thus respond more readily to non-axisymmetric
  perturbations. We also find that the vertical velocity dispersion of outward
  migrators always decreases, whereas the opposite holds for inward
  migrators. To first order, newly arrived migrators simply replace stars that
  have migrated off to other radii, thus inheriting the vertical bias of the
  latter. Extreme migrators might therefore be recognized, if present, by the
  unexpectedly small amplitude of their vertical excursions. Our results show
  that migration, understood as changes in angular momentum that preserve
  circularity, can affect strongly the thin disk, but cast doubts on models
  that envision the Galactic thick disk as a relic of radial migration.
\end{abstract}


\altaffiltext{1}{Department of Astronomy, University of Wisconsin, 2535
  Sterling Hall, 475 N. Charter Street, Madison, WI 53076, USA.
  \href{mailto:ciro@astro.wisc.edu}{e-mail:ciro@astro.wisc.edu}}

\altaffiltext{2}{Department of Physics and
  Astronomy, University of Victoria, Victoria, BC,  V8P 5C2, Canada}

\altaffiltext{3}{Observatorio Astron\'omico, Universidad Nacional de
  C\'ordoba, C\'ordoba X5000BGR, Argentina}

\altaffiltext{4}{Alfred P. Sloan Fellow}

\altaffiltext{5}{Senior CIfAR Fellow}

\section{Introduction}
\label{sec:intro}

Cold, thin stellar disks are dynamically fragile entities that react strongly
to changes in the gravitational potential, such as those that result from
changes in the mass of the galaxy itself, from external perturbations such as
the accretion of a satellite or the tidal effects of substructure, or from
internal processes such as bar instability or the development of spiral
structure.  In the absence of dissipation, these perturbations generally heat
the disk, moving stars away from their birth radii at the expense of
decreasing the circularity of their orbits and increasing the amplitude of
their vertical oscillations. Stars on nearly circular orbits must have
therefore been born at about the same radius where they are found today.

One difficulty with this simple scenario arises in our Galaxy from the weak
correlation between age and metallicity for thin disk stars in the solar
neighborhood \citep{Edvardsson1993}, which is essentially flat for ages
between $0$ and $8$ Gyr \citep[see, e.g.,][and references
therein]{Bergemann2014}. Although this may be taken to suggest a simple (but
somewhat contrived) scenario where the average metallicity of the interstellar
medium (ISM) at the solar circle has not changed since $z\sim 1$, the sizable
scatter in the correlation would also require the ISM to have been rather
in-homogeneous in the past, a possibility that seems at face value unattractive
\citep[see][for a recent review]{Feltzing2013}.

An alternative view holds that thin disk stars do not necessarily remain close
to their birth radii \citep{Wielen1996}. If true, then long-lived stars in the
solar neighborhood might actually sample the chemical properties of the
Galaxy over a wide range of radii at the time of their formation, possibly
blurring any correlation between the age and metallicity of stars selected in
local samples. This scenario gained support after Sellwood and Binney, in an
influential paper, identified a mechanism that allows stars to migrate in
radius whilst remaining on nearly circular orbits \citep{Sellwood2002}. The
mechanism relies on angular momentum exchanges between stars near corotation
of transient spiral patterns, and allows stars to switch radii without
substantially heating the disk. Given that spiral patterns seem ubiquitous in
disk galaxies, this mechanism must operate to some degree in most galaxy
disks.

Since its inception, the corotation resonance (CR) mechanism of
\cite{Sellwood2002} has been the subject of a number of studies in the
literature. These studies have generally confirmed the validity of the
arguments laid out in their original paper and extended their
application to multiple, interacting spiral patterns
\citep{Minchev2010, Brunetti2011}, but have also raised a number of
questions about the initial and final properties of migrating stars
and about the consequences of radial migration for the vertical
structure of the disk. It is unclear, for example, what biases, if
any, might affect the kinematics or spatial distribution of migrators,
what strategies might be used to identify migrators observationally,
or whether radial migration plays a role in the origin of the thick
disk of the Galaxy.

The latter issue, in particular, has been debated vigorously,
especially since \citet{Schonrich2009a} proposed that the chemical
differentiation between thin and thick disks could be explained solely
by radial migration, which might have brought stars with high vertical
kinetic energy from the inner Galaxy out to the solar
neighborhood. \citet{Minchev2012}, on the other hand, have argued that
migration does little for disk thickening because outward migrators
should cool vertically as they move out, a result that has been
further debated by \citet{Solway2012} and \citet{Roskar2013}.

\begin{figure}
  \begin{center}
    \includegraphics[width=0.49\textwidth]{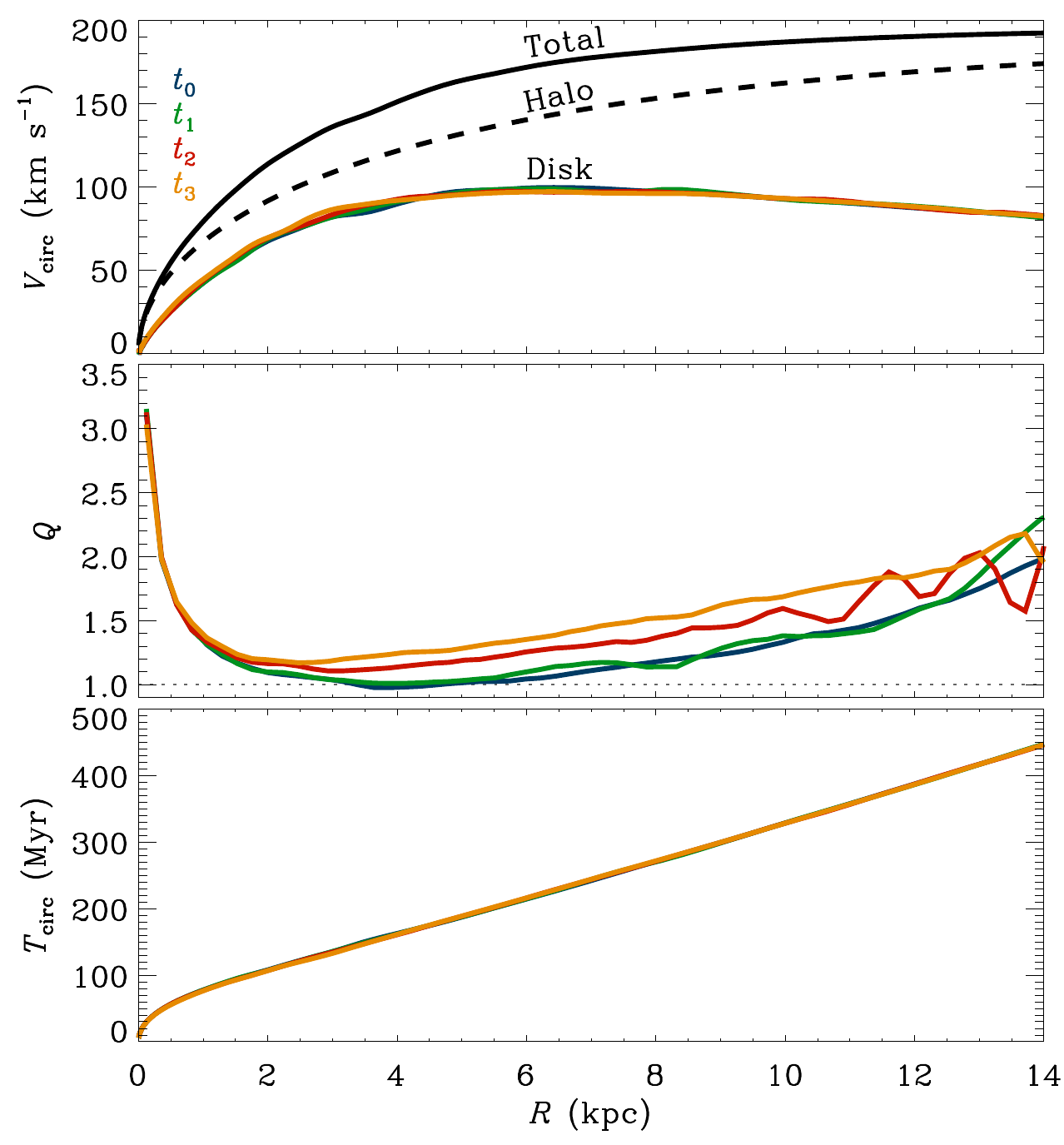}
  \end{center}
  \caption{Properties of the galaxy model. Top: circular velocity,
    $V_{\rm circ}^2 = R\, \partial \Phi/\partial R$, profile for the spiral
    simulation, at various times. No substantial evolution is observed in the
    overall mass distribution, even after $5$ Gyr of evolution. The contribution of
    the disk at various times is shown by the colored curves; that of the
    (static) dark halo is shown by the dashed line; the total is indicated by
    the black solid line. Middle: Toomre's $Q$ parameter profile of the spiral
    disk at different times. Bottom: circular orbit period, $T_{\rm circ} =
    2\pi R / V_{\rm circ}$, as a function of radius at different
    times.}
\label{FigGxModel}
\end{figure}

This unsettled state of affairs might be traced, in part, to the disparate
numerical contexts used to study the effects of migration. Although most
published work has relied on direct numerical simulation, some studies are
based on the response of equilibrium disk models to transient spiral modes
\citep{Sellwood2002,Grand2012,Solway2012,Sellwood2014} or to the tidal
perturbation of a massive satellite
\citep{Kazantzidis2008,Villalobos2008,Bird2012}, whereas others have focused
on hydrodynamical simulations of disks of gas and stars that evolve in
isolation \citep{Minchev2011,Minchev2012} or that result from the dissipative
collapse of a rotating sphere of gas in the gravitational potential of a dark
matter halo \citep{Roskar2008,Roskar2012,Roskar2013,Loebman2011}.

The first approach benefits from the controlled nature of the initial disk,
which allows for the identification of numerical artifact, but suffers from
the arbitrariness involved in choosing the orbit and mass of the 
perturbers, or from differences in the origin of the spiral perturbations,
which either develop freely or result from the forcing of particular modes
with predetermined amplitude and duration. The second approach allows the disk
to evolve more freely, but its analysis is obscured by the complexities that
arise from the continuously evolving disk potential and from the presence of
rapidly evolving multiple modes and instabilities. Furthermore, the numerical
resolution of hydrodynamical simulations is inevitably much lower than that
attainable with $N$-body models, making it much harder to discern and isolate
the effects of noise associated with the finite number of particles used.

Finally, there are also semantic issues, since some authors use the
term ``radial migration'' as a catch-all phrase to denote any radial
displacement of a star, regardless of its cause or effect on its
orbit, whereas others reserve the term ``migration'' for changes in
the guiding center radius (i.e., the orbital angular momentum) of
orbits that remain nearly circular through the process. We shall adopt
the latter meaning in this contribution.

We address some of these issues here using numerical simulations that evolve a
kinematically cold self-gravitating disk of stars in the fixed potential of a
cosmologically motivated dark matter halo. Spiral patterns are seeded by
perturbations modeled on the gravitational influence of molecular clouds and
persist almost unchanged for over $5$ Gyr, the duration of the simulation
(more than $40$ full rotations at the exponential disk scalelength). Our
simulations differ from earlier work mainly on numerical resolution: the disk
is made up of $10^8$ particles, an improvement of nearly two orders of
magnitude over earlier work. The large number of particles allows us to study
the orbital evolution of individual stars without having to worry about the
discreteness noise associated with the particle representation of the disk.

Our simulations also differ from earlier attempts because the disk is not
subject to the rapid development of instabilities that may change its radial
or vertical structure; a control run without perturbers shows that the disk
does not develop strong non-axisymmetric features over several rotation
periods. Finally, our simulations also stand out because the spiral features
seeded by the ``molecular clouds'' do not heat appreciably the disk and
persist for over $5$ Gyr. Both the vertical and azimuthally averaged structure of the
disk remain basically unchanged throughout.

The plan for this paper is as follows. We describe the numerical simulations
used in this work in Section~\ref{SecSims}, present our main results in
Section~\ref{SecRes}, and conclude with a brief summary in
Section~\ref{SecConc}.

\begin{figure}
  \begin{center}
    \includegraphics[width=0.49\textwidth]{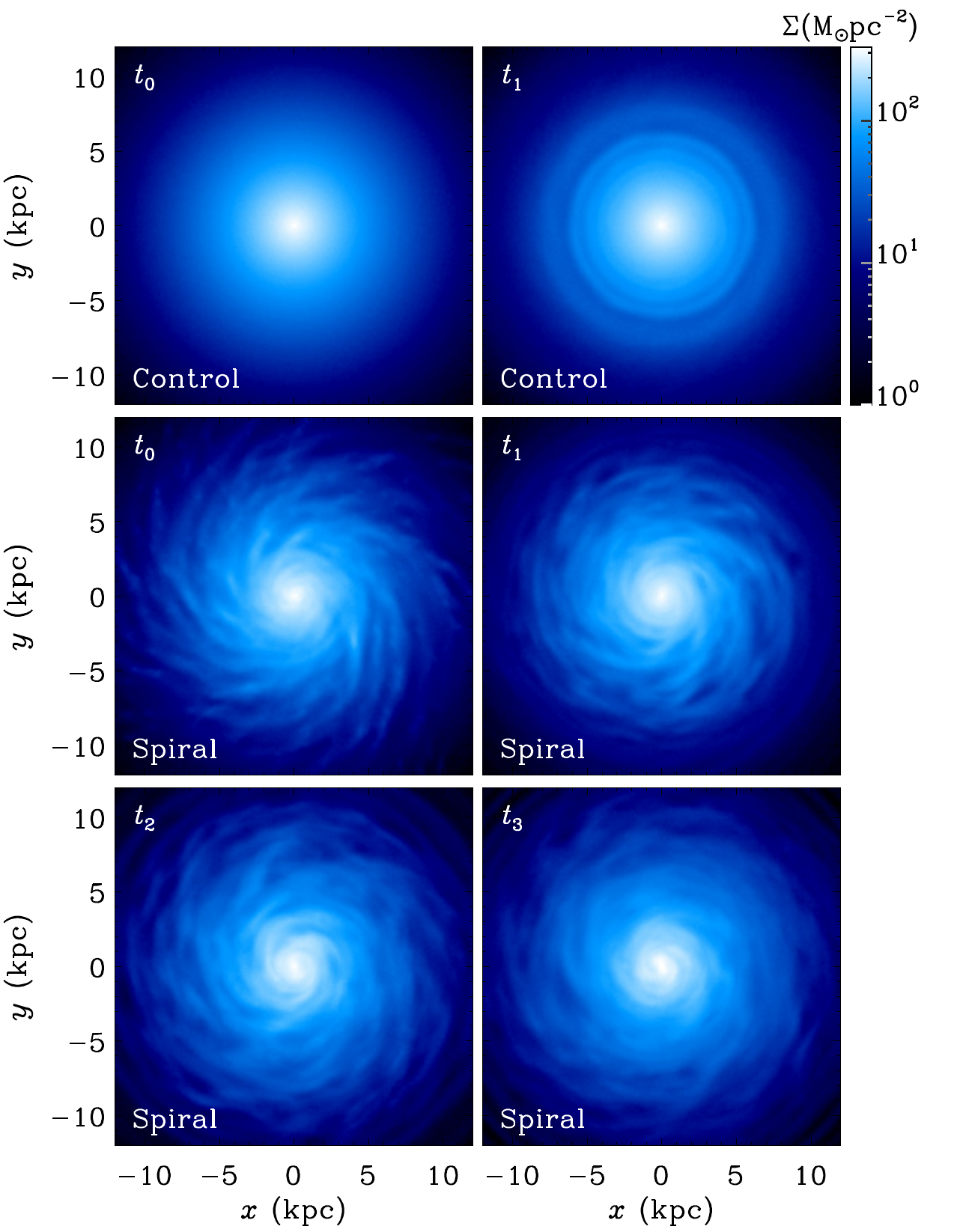}
  \end{center}
  \caption{Face-on projected stellar surface density of the
    simulations at different times, as labeled. The top row shows the
    control simulation; the others correspond to the ``spiral''
    disk. For the spiral disk, $t_0$ corresponds to the time just
    after the molecular-cloud perturbers are removed. Note that at
    that time the spiral arms are already in place. The times shown
    correspond to times measured after $t_0$: $t_1=831$ Myr, $t_2=
    3032$ Myr, and $t_3 = 5477$ Myr.}
\label{FigFaceOn}
\end{figure}

\section{Numerical Methods}
\label{SecSims}

\subsection{The Galaxy Model}

We use here $N$-body simulations of the same galaxy model presented by
\citet{DOnghia2013} . For the sake of completeness we give here a brief
description of the simulations and refer the interested reader to the original
reference for further details.  The models evolve a self-gravitating disk of
particles embedded in a rigid halo potential described by a Hernquist profile
\citep{Hernquist1990} with total mass $M_{\rm halo} = 9.5 \times 10^{11}
\;{\rm M}_{\odot}$ and scale length $r_{\rm halo} = 29.8$ kpc.

The disk is a $10^8$ particle realization of an exponential disk of
constant scaleheight:

\begin{eqnarray}\label{EqDensProf}
  \rho(R,z) &=& \Sigma(R) \zeta(z) \nonumber \\
  &=& \frac{M_{\rm disk}}{4\pi R_{\rm disk}^2
    z_0} \exp\left(-\frac{R}{R_{\rm disk}}\right){\rm sech}^2\left(\frac{z}{z_0}\right),
\end{eqnarray}

\noindent whose positions and velocities were generated using a modified
version of the method presented by \cite{Hernquist1993}, as described by
\citet{Springel2005}. We adopt a radial scalelength $R_{\rm disk} = 3.13$ kpc,
scaleheight $z_{0} = 0.1\, R_{\rm disk} = 0.313$ kpc and total disk mass
$M_{\rm disk} = 1.81 \times 10^{10} \;{\rm M}_{\odot}$.

\begin{figure}
  \begin{center}
    \includegraphics[width=0.49\textwidth]{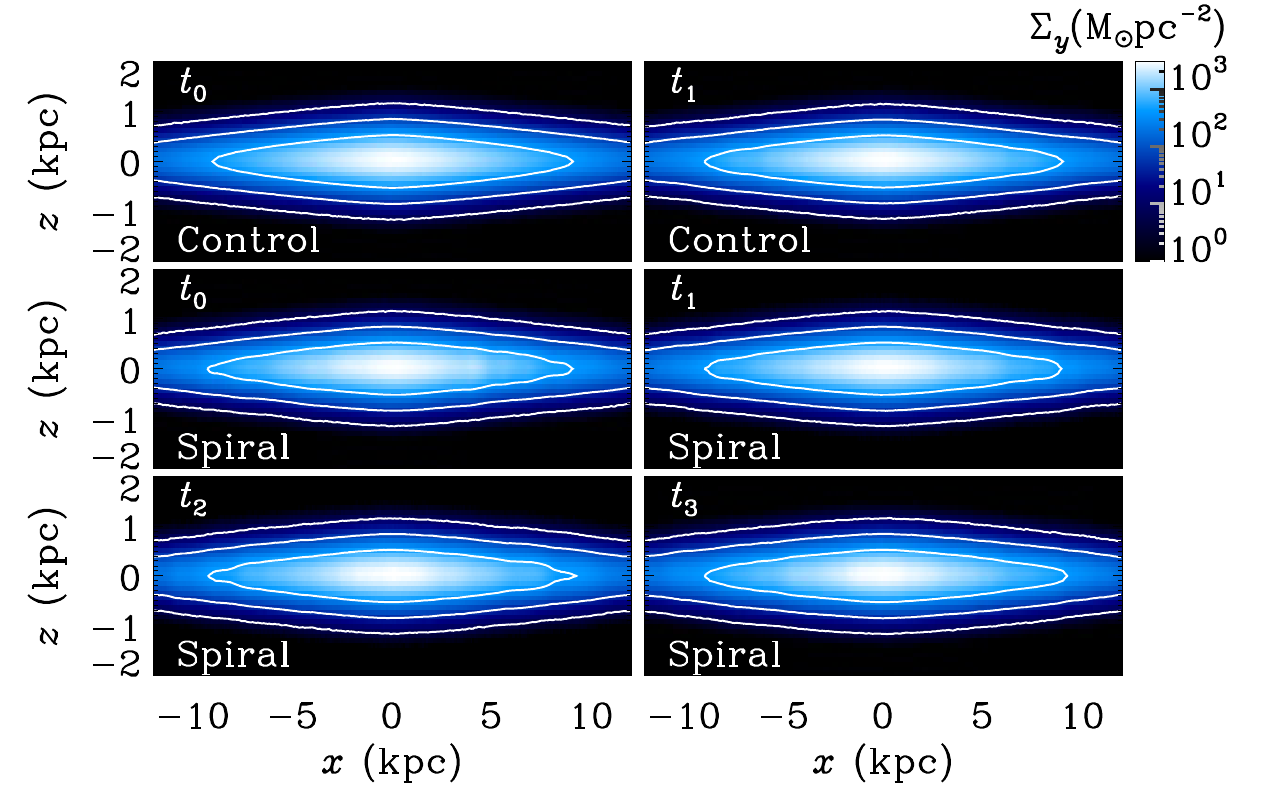}
  \end{center}
  \caption{Same as Figure~\ref{FigFaceOn} but for the edge-on
    projections of the disk.}
\label{FigEdgeOn}
\end{figure}

With these choices, the disk is submaximal, and makes up about 27\% of the
total mass enclosed within two radial scalelengths. A particle in a circular
orbit completes a full rotation at $R_{\rm disk}$ in $136$ Myr. We show in
Figure~\ref{FigGxModel} the circular velocity profile of the galaxy model (top
panel), and the circular orbit timescale, $T_{\rm circ}=2\pi R/V_{\rm
  circ}(R)$, in Myr (bottom panel). The middle panel shows the Toomre
stability parameter $Q=\sigma_R \kappa/(3.36 G \Sigma)$, chosen to ensure the
disk is marginally stable at all radii. (Here $\kappa$ is the epicycle
frequency and $\sigma_R$ the velocity dispersion in radial motions.)

\subsection{The Control Simulation}

We have evolved the galaxy model described in the previous section for
from $t_0=0$ Myr to $t_1=831$ Myr, corresponding to six circular orbit
timescales at $R_{\rm disk}$, or three full rotations at $R=8$ kpc. The
system was evolved using an updated version of \textsc{Gadget}
\citep[see][]{Springel2005b}, using a Plummer-equivalent softening
length for the stellar particles of $5$ pc and default integration
parameters.

After $831$ Myr the ``control'' disk has evolved little from its initial
configuration, as may be seen in the top two panels of Figure~\ref{FigFaceOn},
and has developed no obvious non-axisymmetric features. It does develop,
however, ``ring-like'' features that travel like radial waves through
the disk, and which are easily appreciate in the top right panel of
Figure~\ref{FigFaceOn}.

These waves do not seem to be caused by discreteness effects, since
they appear spontaneously and are coherent over the whole disk. The
radial waves seem to travel back and forth through the disk, without
causing any discernable change in the radial distribution of disk
particles. They do not perturb appreciably the vertical structure of
the disk, as shown in the top two panels of Figure~\ref{FigEdgeOn}, nor
change the surface density profile of the disk, as shown in the left
panel of Figure~\ref{FigSurfDensProf}.

The left panels of Figure~\ref{FigVelDisp} show that the radial waves have
little effect on the kinematics of disk particles: there is little evolution
in the velocity dispersion profiles of the disk over the $831$ Myr that the
disk runs unperturbed. We conclude that our galaxy disk model is not subject
to the strong non-axisymmetric instabilities that have plagued $N$-body studies
of disk galaxies over the years. Although our disks are fully
self-gravitating, it appears as if the number of particles used is sufficient
to suppress discreteness noise and to prevent, for several rotation periods,
the development of strong non-axisymmetric features.

\subsection{The ``Spiral'' Simulation}

As shown by \citet{DOnghia2013}, non-axisymmetric features do develop when the
disk is perturbed in a manner consistent with the presence of $\sim 1000$
``molecular clouds''. We refer the reader to \citet{DOnghia2013} for details
on the numerical implementation but, in brief, the perturbations are modeled
by the addition of softened, $9.5\times 10^{5} \, M_\odot$ particles
constrained to travel in circular orbits and scattered about the disk. The
perturbation is transient, and the molecular-cloud perturbers are removed
$245$ Myr after their inclusion.

The disk reacts strongly to these perturbations, and develops a long-lived
multi-arm spiral pattern, as shown in the bottom four panels of
Figure~\ref{FigFaceOn}. In this sequence, $t_0$ shows the disk just after the
perturbers are removed; $t_1$ corresponds, as in the control simulation, to
$831$ Myr later; whereas $t_2$ and $t_3$ denote later stages of the evolution,
$3032$ Myr and $5477$ Myr after the removal of the perturbers, respectively.

\begin{figure}
  \begin{center}
    \includegraphics[width=0.49\textwidth]{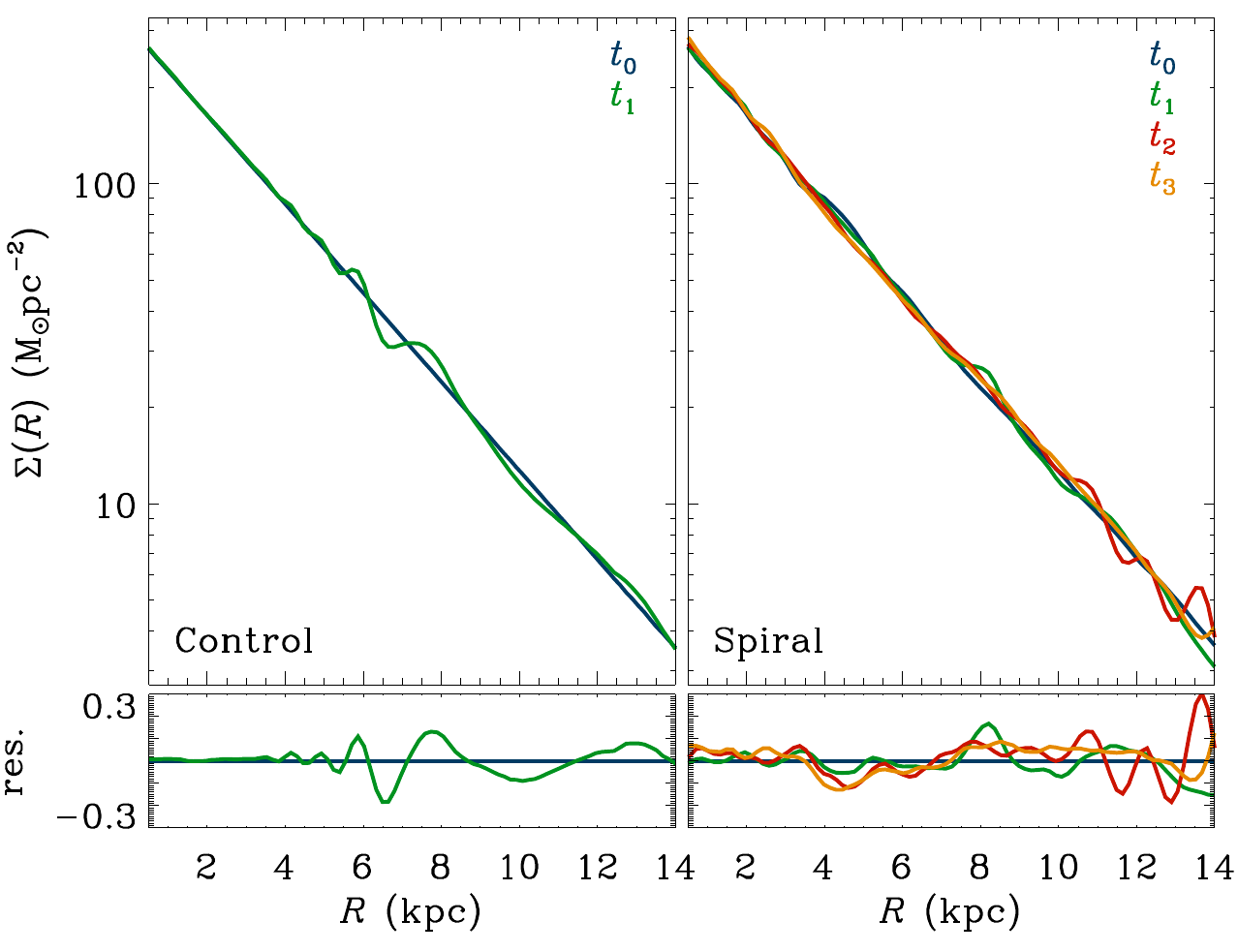}
  \end{center}
  \caption{Surface density profile for the control simulation (left) and the
    spiral simulation (right). Colors correspond to different times, as
    labeled. The residuals for the density are calculated as $\ln \Sigma(R,
    t)/ \Sigma(R,t_0)$ and are shown in the bottom panels. With the exception
    of radial density waves probably associated with small inaccuracies in the
    generation of initial conditions, no systematic change in the exponential
    shape of these profiles is observed.}
\label{FigSurfDensProf}
\end{figure}

Visual inspection of Figure~\ref{FigFaceOn} shows that the multi-arm spirals
seeded by the molecular cloud perturbers are long lived and that they change
little in nature over $\sim 5$ Gyr of evolution. Despite their longevity, the
spiral patterns seem to have little effect on the vertical structure of the
disk, as shown by the bottom four panels of Figure~\ref{FigEdgeOn}. The radial
structure of the disk also remains almost unchanged, as indicated by the
azimuthally averaged surface density profiles at $t=t_0$, $t_1$, $t_2$, and
$t_3$ shown in the right-hand panel of Figure~\ref{FigSurfDensProf}.

Only the in-plane kinematics of the disk seems to be noticeably affected, as
shown by the slight, but monotonic, heating in the radial and azimuthal
directions seen in the right-hand panels of Figure~\ref{FigVelDisp}. The
vertical structure of the disk, however, is unaffected throughout: the
vertical velocity dispersion profile, $\sigma_z(R)$, at $t_3$ is
indistinguishable from that at $t=t_0$ (see bottom right panel of
Figure~\ref{FigVelDisp}).

\subsection{Azimuthal Structure}

We examine the azimuthal structure of the disks in Figure~\ref{FigPotSigma},
where we show the fluctuations in the surface density and in the gravitational
potential at various times as a function of the azimuthal angle $\theta$ and
of the cylindrical radius, $R$.

\begin{figure}
  \begin{center}
    \includegraphics[width=0.49\textwidth]{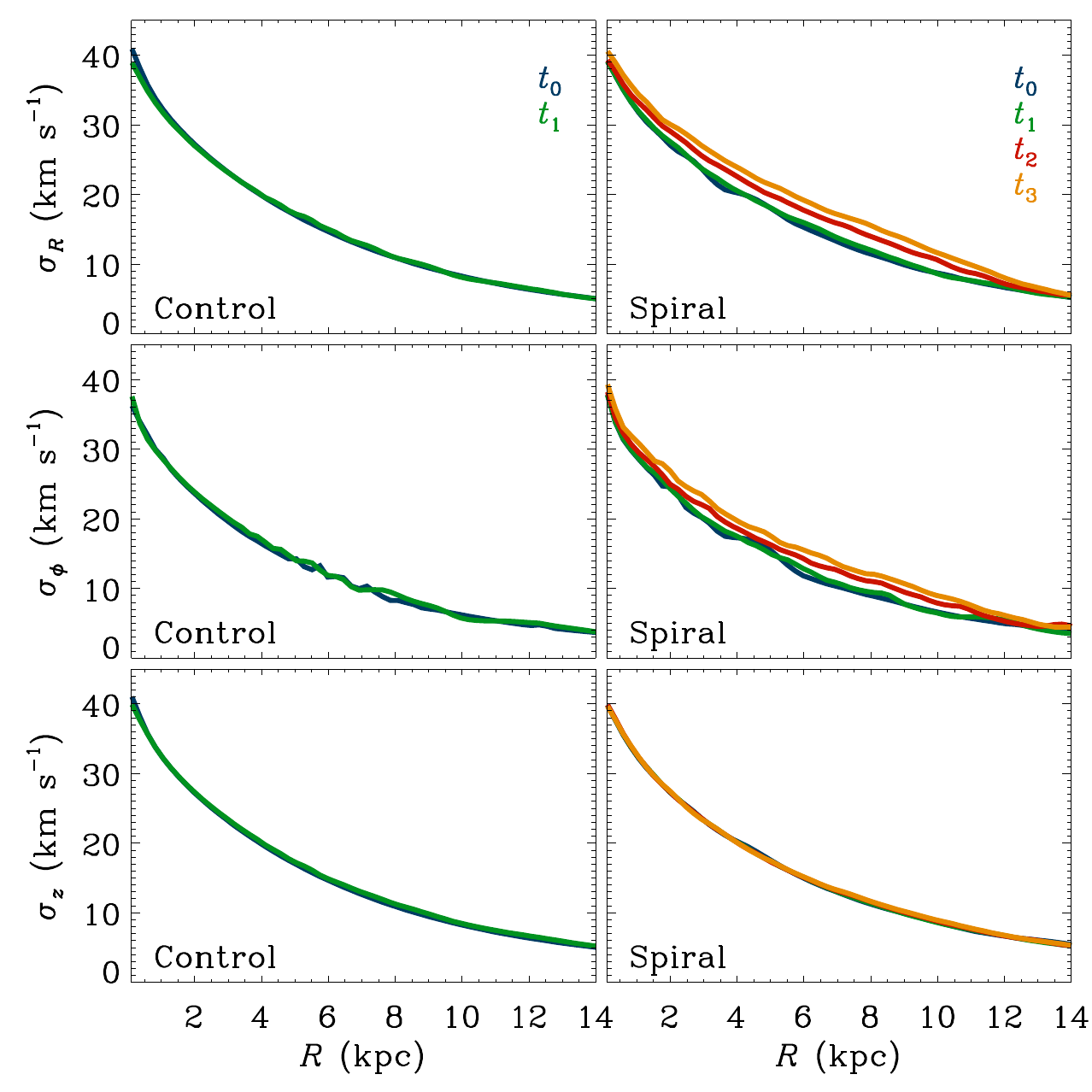}
  \end{center}
  \caption{Radial (top), azimuthal (middle) and vertical (bottom) velocity
    dispersion profiles as a function of radius, for the control simulation
    (left) and the spiral simulation (right). Different colors correspond to
    different times, as labeled. The vertical component $\sigma_z$ is
    remarkably constant even after $\sim 5$ Gyr of evolution, with changes
    smaller than $\sim 3\%$. On the other hand, the in-plane components
    increase slightly but steadily a function of time. The most affected is
    $\sigma_R$, which increases by $\sim 30\%$ at $R=9$ kpc at the final time.}
\label{FigVelDisp}
\end{figure}

\begin{figure*}
  \begin{center}
    \includegraphics[width=0.95\textwidth]{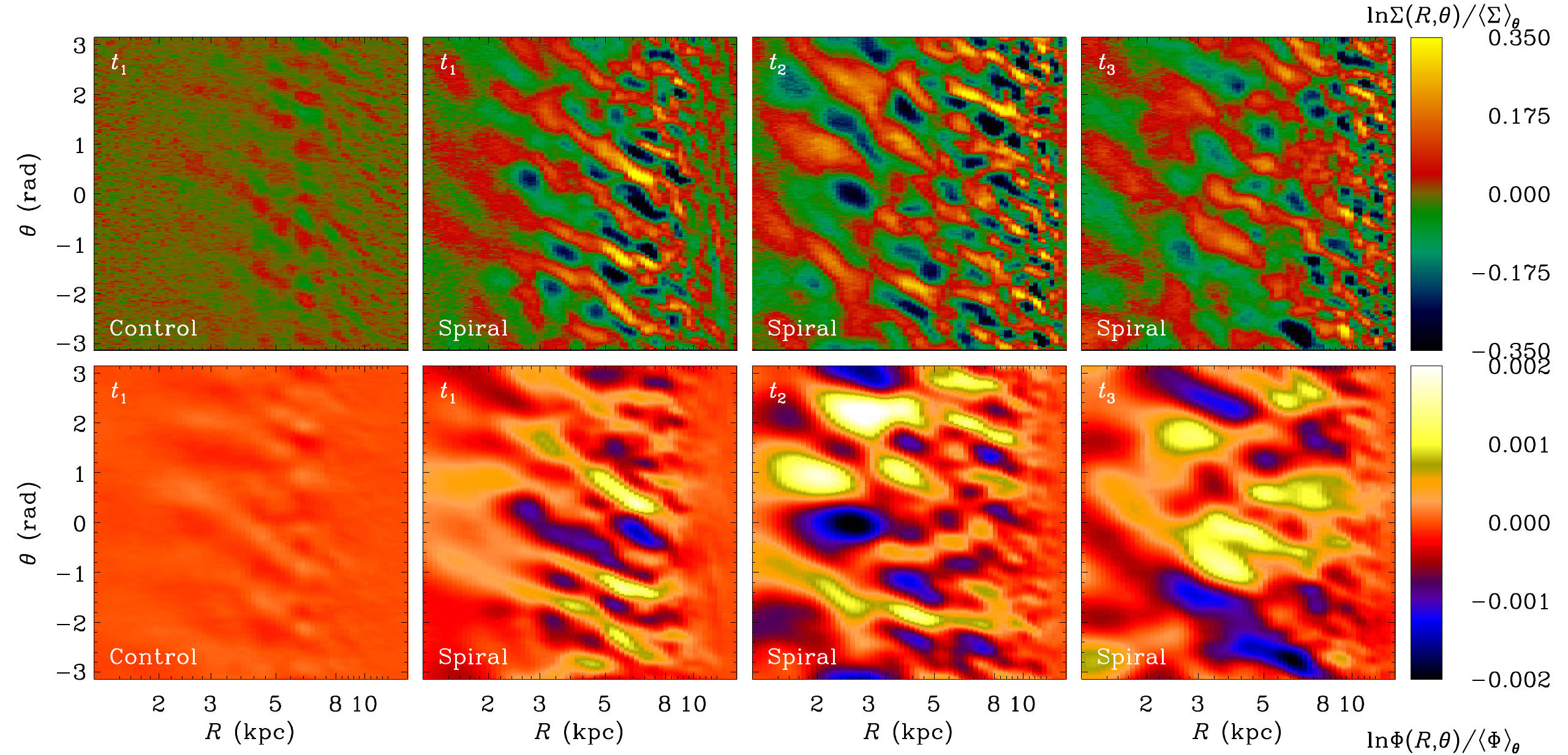}
  \end{center}
  \caption{Fluctuations in the surface density (top row) and the gravitational
    potential at various times for the control simulation (leftmost column)
    and for the spiral simulation. The residuals are computed relative to the
    azimuthally averaged value at given radius as $\ln \Sigma(R,\theta) /
    \langle \Sigma \rangle_\theta$ and $\ln \Phi(R,\theta) / \langle \Phi
    \rangle_\theta$. The interarm fluctuations due to the spiral patterns is
    roughly $\sim 30\%$ in the spiral simulation, but only $\sim 0.1\%$ in the
    potential.}
\label{FigPotSigma}
\end{figure*}

The leftmost panels of Figure~\ref{FigPotSigma} shows that a barely
discernible multi-arm spiral pattern develops in the control
simulation after $831$ Myr of evolution. Given its small amplitude
(the rms fluctuations in the surface density are just $5\%$ at
$R=5$ kpc) this pattern is only evident in the residuals and, as we
will see below, leads to basically no secular evolution in the disk
over the $831$ Myr it was evolved for.  This confirms our earlier
statement that the model galaxy is free from the violent instabilities
or rapidly-growing non-axisymmetric disturbances characteristic of
models where either the galaxy is initially out of equilibrium, or
where discreteness noise seeds perturbations that transform the galaxy
over a short period of time.

The multi-arm spiral structure of the ``spiral'' simulation stands out
clearly in Figure~\ref{FigPotSigma}, with interarm surface density
fluctuations of roughly $\pm 30\%$. The actual gravitational potential
fluctuations are much smaller, however, due to the presence of the
dark matter halo. The spiral structure in the ``spiral'' simulation
remains roughly constant from $t_0$ to $t_3$, and is characterized by
a radially increasing number of arms that bifurcate outward: at $R=2$
kpc, near the center, only about $4$ arms are clearly defined, whereas
more than $8$ arms can be counted $10$ kpc away from the center. We
ascribe the very slight evolution in the spiral structure from $t_0$
to $t_3$ to the weak, but monotonic, radial heating of the disk seen
in the increase of the radial velocity dispersion shown in
Figure~\ref{FigVelDisp}. We emphasize again that the overall heating
is quite modest; in $5$ Gyr of evolution the radial velocity
dispersion at $R=5$ kpc increases from $18$ km s\textsuperscript{-1}
to just over $21$ km s\textsuperscript{-1}. The increase is even
smaller at other radii.

\begin{figure}
  \begin{center}
    \includegraphics[width=0.49\textwidth]{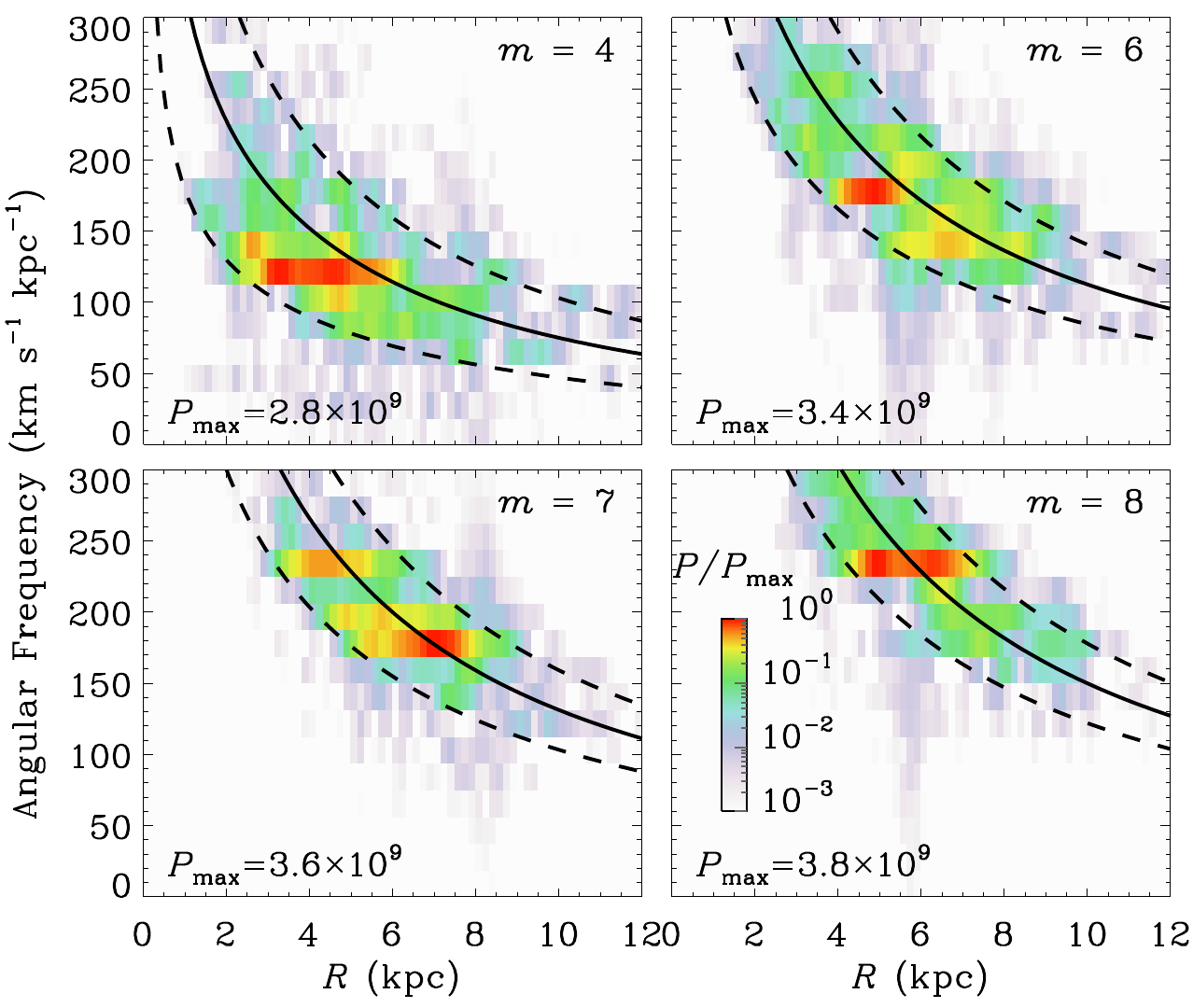}
  \end{center}
  \caption{Power as a function of radius and frequency for different Fourier
    harmonics, $m$, in the spiral disk. The spectrogram is computed using
    snapshots spaced by $5$ Myr in the time interval $930<t/$Myr$<1260$. In
    each panel, the solid line shows $m$ times the circular angular frequency,
    $\Omega_{\rm circ}=V_{\rm circ}/R$. The dashed lines correspond to the
    Lindblad resonances, $m\, \Omega_{\rm circ} \pm \kappa$, where $\kappa$ is
    the usual epicycle frequency. Note the presence of persistent multi-arm
    wave-like perturbations that travel at roughly the circular orbit speed at
    the radius where they dominate. }
\label{FigSpGram}
\end{figure}

The spirals shown in Figure~\ref{FigPotSigma} have pattern speeds that
approximately match the angular speed of the disk at the radii where
they are most prominent. This is seen in the spectrograms of
Figure~\ref{FigSpGram}, where we show the power spectra for the $m=4$,
$6$, $7$, and $8$ harmonic modes, measured from snapshots sampled
every $5$ Myr for the spiral disk in the $930<t/$Myr$<1260$
interval. Note that each mode spans a limited radial range and that
the higher modes travel with a pattern speed similar to that of
circular orbits at that radius.

As discussed by \citet{DOnghia2013}, arms in the pattern fade away and
recur while traveling at about the circular angular frequency, in a
seemingly ``self-perpetuating'' cycle. This implies that stars on
nearly circular orbits will be consistently near CR of the spiral mode
that dominates at each radius, favoring the corotation scattering
mechanism of \citet{Sellwood2002}. We examine its effect over many
rotation periods next.

\section{Results}
\label{SecRes}

\subsection{Radial Migration}
\label{SecRadMig}

In a stationary potential and in the absence of non-axisymmetric perturbations
disk stars should conserve both their energy, $E$, and the $z$-component of
their specific angular momentum, $J_z$. The presence of the transient spiral
patterns discussed in the last section, however, leads to exchanges of angular
momentum amongst disk particles and to variations in their orbital energies.

\begin{figure}
  \begin{center}
    \includegraphics[width=0.49\textwidth]{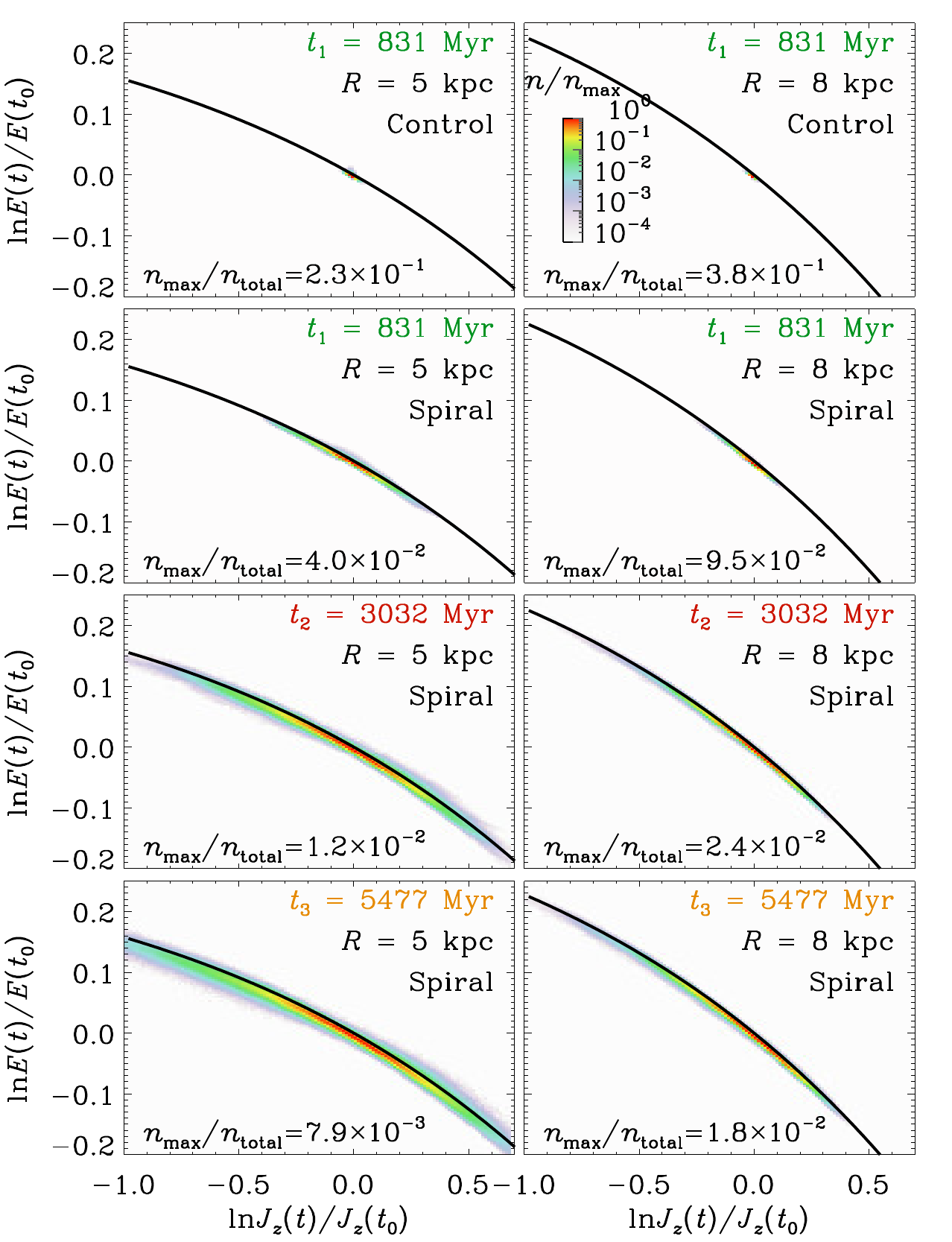}
  \end{center}
  \caption{Changes in energy, $E$, and of orbital angular momentum $J_z$
    measured for particles identified at $t_0$ at two different radii:
    $R=5\pm0.25$ kpc (left) and $R=8\pm 0.25$ kpc (right). The top row show
    the control simulation; the rest correspond to the spiral disk, at the
    times labeled in each panel. The solid line shows the changes in $E$ and
    $J_z$ needed for a particles to remain on a circular orbit. Stars
      are binned in a $128\times 128$ grid, with bin sizes of $\delta \ln E =
      3.5\times 10^{-3}$ and $\delta \ln J_z = 1.3\times 10^{-2}$. The color
      scale encodes the number of points in each pixel $n$ normalized by
      the maximum $n_{\rm max}$.}
\label{FigdEdJz}
\end{figure}

\begin{figure}
  \begin{center}
   \includegraphics[width=0.49\textwidth]{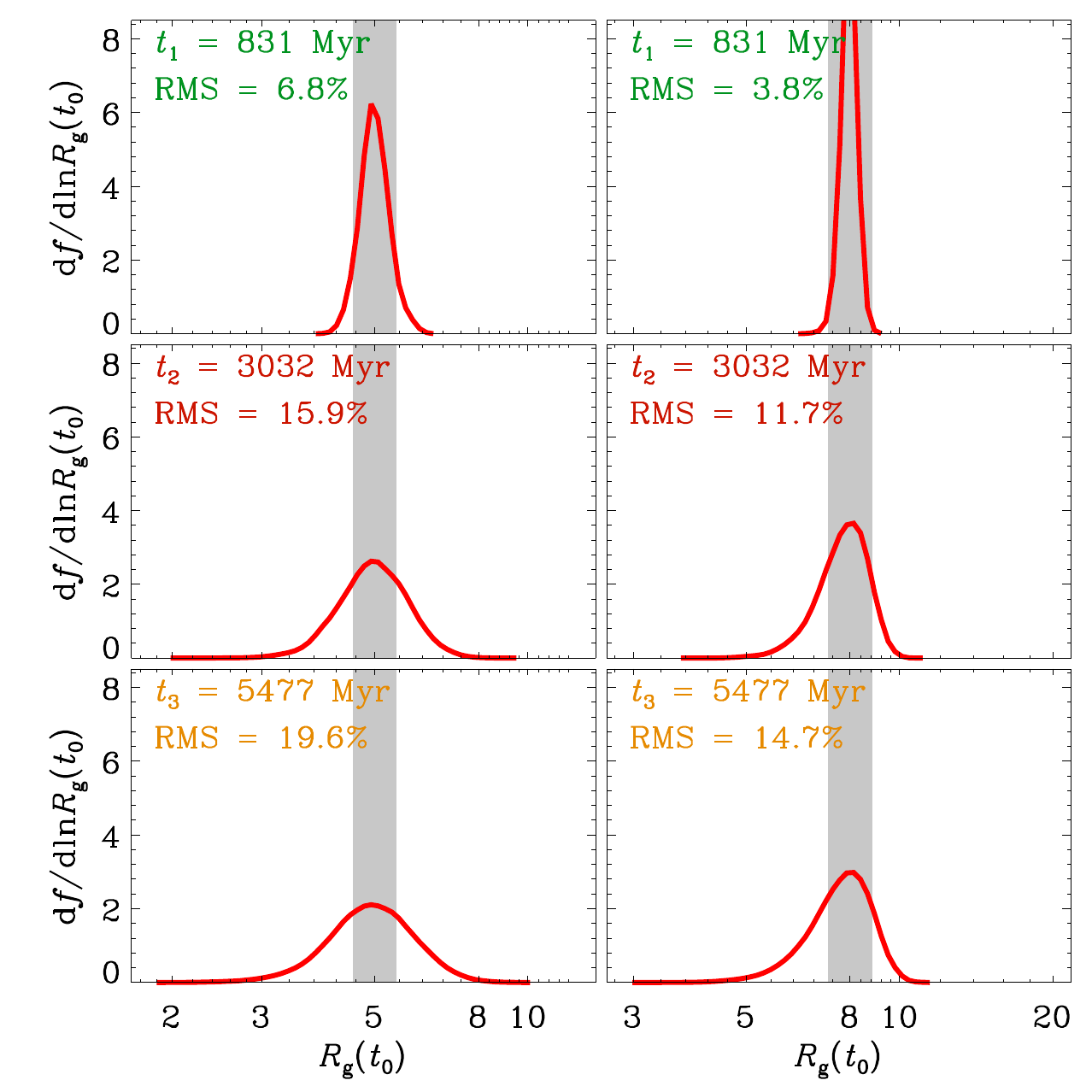}
  \end{center}
  \caption{Distribution, at $t_0$, of guiding center radii of particles that
    migrate, by $t_1$ (top), $t_2$ (middle) and $t_3$ (bottom), to
    $R_g=5\pm0.5$ kpc (left column) or $R_g=8\pm0.8$ kpc. Labels in each panel
    indicate the rms of the distribution, which increases roughly like
    $t^{1/2}$. }
\label{FigHistdR}
\end{figure}

We examine these changes by selecting, at $t_0$, particles within two
$500$ pc wide cylindrical shells centered at $R=5$ and $8$ kpc,
respectively. We track these particles at later times and compare their
current energies and angular momenta with those at $t_0$. The comparison is
shown in Figure~\ref{FigdEdJz} for the control simulation (at $t_1$) and for the
spiral simulation at $t_1$, $t_2$ and $t_3$.

Particles in the control simulation exhibit very slight changes in their
values of $E$ and $J_z$, as expected given the smooth radial and azimuthal
structure of the control disk at all times during its evolution.

Particles in the spiral simulation, on the other hand, undergo correlated
changes in $E$ and $J_z$ that can reach (and exceed, in some cases) a factor
of $\sim 2$ in $J_z$. Recalling that the guiding center radius, $R_{\rm g}$, of a
star in a nearly circular orbit is directly proportional to $J_z$;
\begin{equation}
 J_z(R_{\rm g})  = R_{\rm g} \, V_{\rm circ}(R_{\rm g}),
\end{equation}
(where $V_{\rm circ}^2(R) = R\, \partial \Phi/\partial R$ and $\Phi$ is the
gravitational potential on the disk plane) this implies changes in the guiding
center that can exceed a factor of $\sim 2$.

These variations, however, do not alter the nearly circular nature of the
orbits. Indeed, the correlated changes in $E$ and $J_z$ keep the stars very
close to the minimum energy consistent with its angular momentum (indicated by
the solid curves in Figure~\ref{FigdEdJz}), which defines circular orbits. Thus,
the spiral perturbations shown in Figs.~\ref{FigFaceOn} and ~\ref{FigPotSigma}
cause substantial migration in the spiral disk with minimal heating, as
envisioned in the CR mechanism of \citet{Sellwood2002}.

Figure~\ref{FigdEdJz} also hints that, as time goes by, the radial displacements
of disk particles increase monotonically. We show this more explicitly in
Figure~\ref{FigHistdR}, where we plot the distribution of guiding center radii
at $t=t_0$ of particles that migrate at later times to $R_{\rm g}(t)=5 \pm0.5$
kpc (left panels)
or $8\pm0.8$ kpc (right panels).  The width of these distributions increase with time
roughly as $t^{1/2}$, suggestive of a simple diffusion mechanism.

After more than $5$ Gyr of evolution (i.e., at $t_3$) more than $10\%$ of
particles with $R_{\rm g}=5\pm 0.5$ kpc have migrated to their current orbits from
radii that were initially either $40\%$ smaller or larger. At $8$ kpc the
fraction of such extreme migrators reduces to $6\%$ in the same time, presumably
because of the longer orbital times characteristic of larger radii. The radial
distribution of migrators at $8$ kpc is also clearly asymmetric, with a larger
fraction of extreme migrators that come from the inner regions. This is again
not surprising, since there are few particles beyond $8$ kpc in the disk.

\begin{figure}
  \begin{center}
    \includegraphics[width=0.49\textwidth]{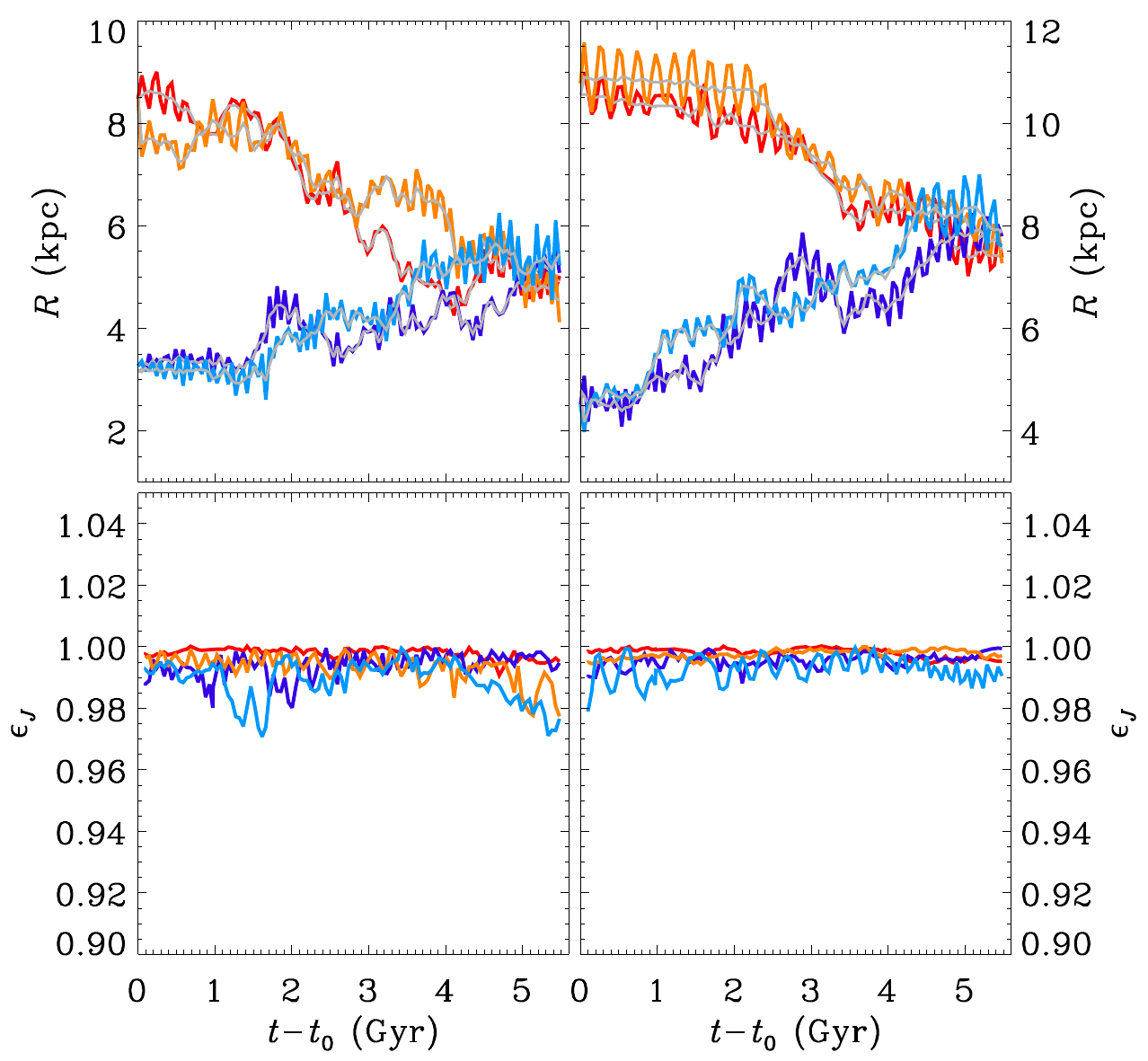}
  \end{center}
  \caption{Top: orbits of extreme migrators that end up at $R\approx 5$ kpc
    (left panel) and at $R=8$ kpc (right panel). Two outward and two inward
    migrators are shown  in each panel. In each case the gray line shows the
    instantaneous guiding center of the orbit. The bottom panels show the
    evolution of the circularity parameter, $\epsilon_J=J_z/J_{\rm circ}(E)$,
    of the orbit of each migrator shown in the top panels. Note that the
    orbits of the migrators remain nearly circular despite the large changes
    in orbital angular momentum.}
\label{FigExtMig}
\end{figure}

We show in Figure~\ref{FigExtMig} the evolution in radius of a few ``extreme''
migrators that have moved roughly $\sim 50\%$ inward or outward to reach
their final radius. The top left and right panels correspond to migrators that
reach $R_{\rm g}\sim 5$ or $8$ kpc at $t_3$, respectively.  In spite of the large
changes in guiding center, the migrators' orbital circularities barely
change. This is shown in the bottom panels of Figure~\ref{FigExtMig}, where we
plot the evolution of the circularity parameter, $\epsilon_J=J_z/J_{\rm
  circ}(E)$, defined as the specific angular momentum in units of that of a
circular orbit of the same energy. As these panels show, even extreme
migrators stay on nearly circular orbits throughout the evolution.

\subsection{Migrator Properties}
\label{SecPropMig}

We explore the properties of migrators in Figure~\ref{FigVelDispMig}, where we
show, in the left-hand panels, the vertical velocity dispersion of particles
with guiding center radii $R_{\rm g}(t)=5\pm 0.5$ kpc (shaded area), as a function of
their original guiding center, $R_{\rm g}(t_0)$. Particles to the left of the shaded
area have migrated outwards to reach $\sim 5$ kpc; those to the right have
migrated inward.  For reference, the thick dashed curves in each panel of this
figure indicate the vertical velocity dispersion profile of all particles in
the disk, $\sigma_z(R)$.

The blue curves in Figure~\ref{FigVelDispMig} show the current velocity
dispersion of subsets of particles binned according to $R_{\rm g}(t_0)$. The
inverted ``U'' shape indicates that migrators are a kinematically biased
population with lower velocity dispersions than non-migrators, a bias that
increases the farther away a particle has migrated from in order to reach its
current radius. The bias affects both inward and outward migrators, and holds
at all times. The bias is not only kinematical; as expected given their low
vertical velocity dispersion, the vertical distribution of migrators is also
thinner than that of non-migrators, as shown by their current vertical
scaleheight (see the blue curves in the right-hand panels of
Figure~\ref{FigVelDispMig}). Again, the farther particles have migrated from the
thinner their current vertical distribution.

The cause of this vertical bias is twofold.  Firstly, migrating stars are a
heavily-biased subset of particles at their original radii, a ``provenance
bias'' that increases the farther a particle migrates in a given time. This
may be seen by comparing the red curves in Figure ~\ref{FigVelDispMig}, which
show the velocity dispersion and scaleheight of migrators at their initial
radius, with the dashed black line, which shows the same but averaged for all
stars at that radius. The same vertical bias is clearly present at their
radius of origin: {\it radial migration mainly affects thinner-than-average
  populations at both their destination and provenance radii.}

A second reason for the migrators' vertical bias may be seen by comparing the
initial and final velocity dispersions of migrating stars. This is given by
the difference, at given $R_{\rm g}(t_0)$, between the red and blue curves in
Figure~\ref{FigVelDispMig}. Outward migrators (to the left of the gray zone)
clearly cool as they move out (i.e., the blue curve is below the red), whereas
the opposite holds for inward migrators. We note that this in itself would not
be enough to explain the vertical bias at the final radius, since outward
(inward) migrators originate in regions of higher (lower) velocity dispersion
of the disk. However, Figure~\ref{FigVelDispMig} shows that such cooling or
heating is but a small correction to the ``provenance bias'' affecting all
migrators.

We have checked that a similar provenance bias and heating/cooling process
affect migrators at other radii, and that similar results are obtained when
analyzing the other two components of the velocity dispersion, $\sigma_R$ and
$\sigma_\phi$. Therefore, radial migration does not only operate most
effectively on particles closest to the disk, but also on those closest to
circular orbits. These biases have been hinted at in earlier work \citep[see,
e.g.,][]{Minchev2012,Solway2012,Bird2012,Roskar2013} but their importance has
not been properly appreciated, obscured perhaps by the concurrence of other
heating mechanisms (i.e., satellite perturbations, disk growth, major disk
instabilities) whose effectiveness in displacing stars radially can exceed
that of the corotation scattering mechanism we focus on here
\citep[although see the case of the low surface brightness disk galaxy
  model explored in][]{Loebman2011}.

With the benefit of hindsight, our results are not surprising. Radial
migration is caused by non-axisymmetric patterns in the disk that are more
readily felt by stars that remain closer to the disk during their orbits;
thus, the smaller the vertical amplitude of their oscillations the farther
stars can migrate. This is, of course, true at all radii affected by the
spiral patterns. If radial migration is not to cause vertical heating or a
change in the surface density profile (as is the case in our simulation) then
migrating stars at their final location must simply ``substitute'' other stars
that have migrated off, which were themselves biased in the same way. In other
words, to first order radial migration leads stars to merely ``exchange''
places in the disk, and affects first and foremost stars in the very thin
disk.

\subsection{Effects of Radial Migration}
\label{SecEffMig}

Our results extend and clarify earlier work on the possible effects of radial
migration. For example, some authors have argued that outward migration causes
the disk to thicken because it would shift stars with high vertical kinetic
energy to regions where the disk gravity is lessened, an mechanism that
\citet{Schonrich2009a} have cited as a plausible origin for the Galactic thick
disk. This assumption has been qualified and contested in later work, which
have argued that migrating stars should cool vertically if they migrate out
and heat up if they migrate inward. This was first proposed by
\citet{Minchev2012c}, who argued that such kinematic changes might signal the
conservation of the vertical action, a result that has been further elaborated
in more recent work
\citep{Solway2012,Minchev2012,Roskar2013}. 

Our results help clarify this issue.  Although it is true that
migrators heat or cool as they move in or out of the disk, these
changes play only a small role in determining the final vertical
distribution of migrators. This is mainly set by the provenance bias
we describe above, which has not been fully accounted for in earlier
work. Because the bias adds to the cooling associated with outward
migration, our results suggest that ``hick disks'' might not result
from the only effect of radial migration, at least in multi-armed
spiral galaxies.

Of course, this does not rule out the possibility that thick disk-stars in the
solar neighborhood might have formed elsewhere in the Galaxy: an external
origin \citep{Abadi2003} or a pre-existing thin disk heated by external
perturbations \citep{Quinn1993}, early mergers \citep{Brook2004}, or the
deepening potential of the forming Galaxy, remain viable options for the
origin of the Galactic thick disk.

Our results also suggest that radial migration could have had a
substantial effect on the thin disk and that, in local volumes such as
the solar neighborhood, the most extreme migrators might actually be
found amongst stars closest to the disk. This is consistent with
observations that suggest that some of the clearest signs of migration
are found amongst thin disk stars.

\begin{figure}
  \begin{center}
    \includegraphics[width=0.49\textwidth]{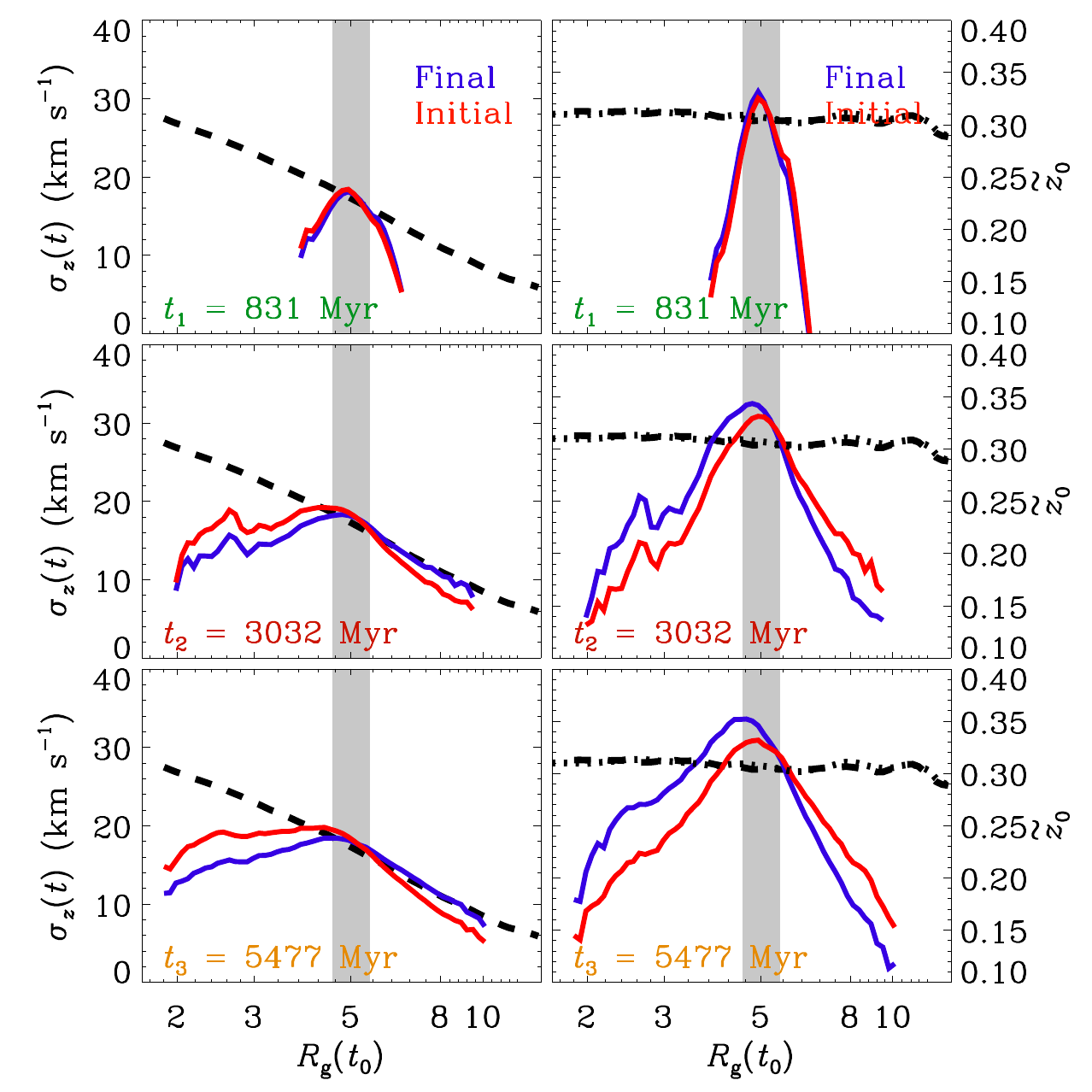}
  \end{center}
  \caption{Left column: vertical velocity dispersion of particles that migrate
    to $R_{\rm g}(t_3) = 5\pm0.5$ kpc, shown as a function of their initial guiding
    center radius, $R_{\rm g}(t_0)$. The velocity dispersion profile of the whole
    disk is shown by the dashed black lines. Blue corresponds to velocity
    dispersions at $t_3$; red indicates $\sigma_z$ at $t_0$. Right column:
    same as left panels, but for the scale-height, defined as $\widetilde{z}_0
    = (2\sqrt{3}/\pi)\langle z^2 \rangle^{1/2}$, and normalized so that
    $\widetilde{z}_0 = z_0$ for a vertical profile of the form $\zeta(z) = 1/2
    \, {\rm sech}^2z/z_0$ (see Eq.~\eqref{EqDensProf}).}
  \label{FigVelDispMig}
\end{figure}

One compelling example involves the origin of the large spread in metallicity
of the thin disk, which spans the range $-0.7<$[Fe/H]$<0.3$ \citep[see,
e.g.,][and references therein]{Navarro2011}. If the spread is caused by radial
migration of stars from the inner and outer Galaxy, then one would expect that
those most chemically distinct should correspond, on average, to the most
extreme migrators. 

As discussed, for example, by \citet{Haywood2008}, this might explain
the puzzling anti-correlation between metallicity and rotation
velocity found for stars in the thin disk. In this scenario, the
anti-correlation merely reflects the difference in angular momentum
expected for inward and outward migrators that have reached the solar
neighborhood. Arrivals from the outer disk (presumably those
populating the metal-poor tail of the thin disk) should have
higher-than-average angular momentum, while the opposite should hold
for those coming from the inner disk. The net effect is a negative
gradient in the $V_\phi$ versus [Fe/H] relation (or, equivalently, in
$R_{\rm g}$ versus [Fe/H]), which is indeed observed
\citep{Lee2011,Haywood2013,Recio-Blanco2014}.

\begin{figure}
  \begin{center}
    \includegraphics[width=0.49\textwidth]{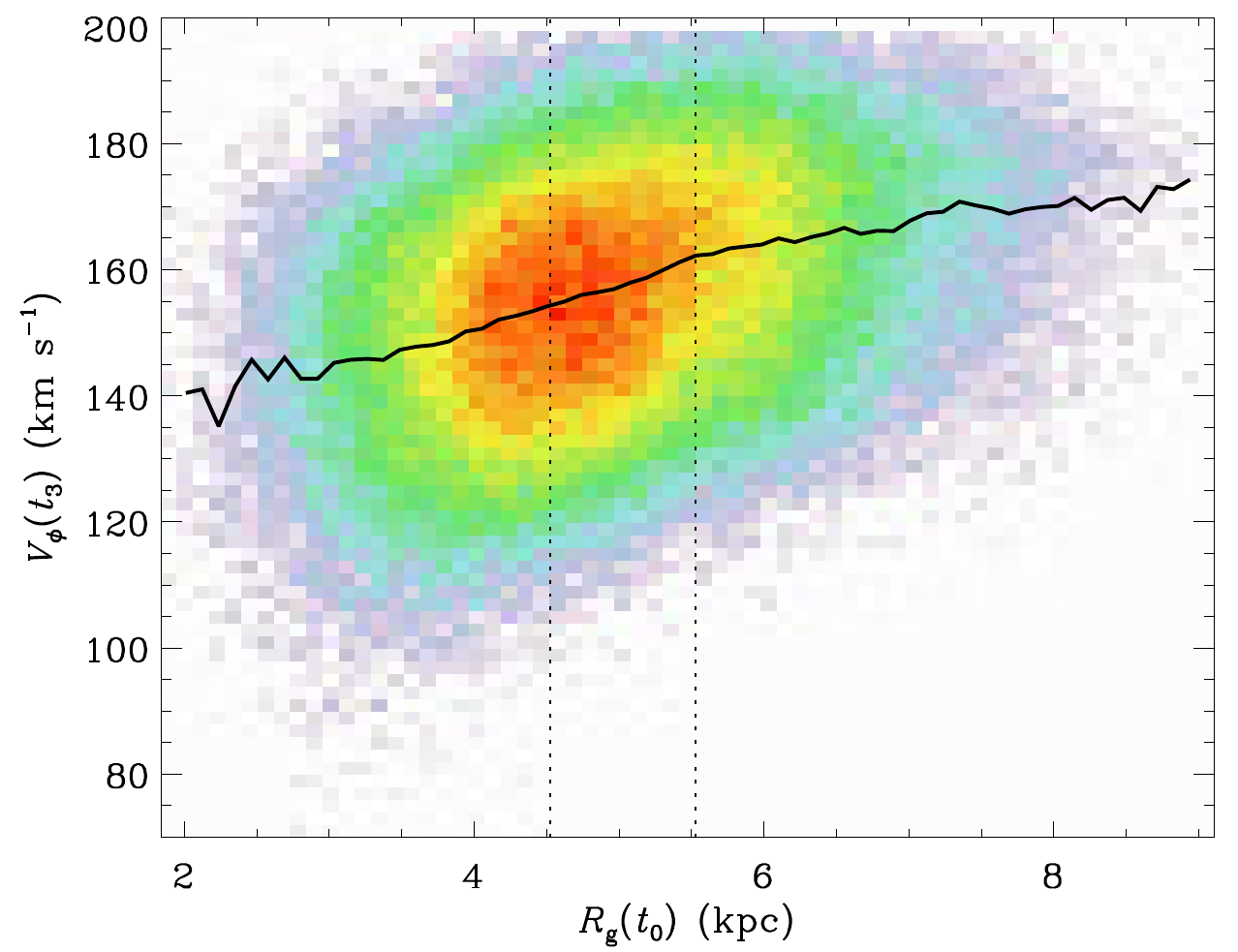}
  \end{center}
  \caption{Azimuthal velocity versus initial guiding center radius for particles
    identified, at the final time $t_3$, within a cylindrical shell of radius
    $R=5\pm 0.5$ kpc. All of these stars have, at $t_3$, guiding center radii in
    the range $3.3 <R_{\rm g}/$kpc$<6.2$. Note that inward migrators from the outer
    disk have higher rotation velocities than outward migrators from the inner
    galaxy.}
\label{FigRg0Vphi}
\end{figure}

Our results confirm this expectation. Figure~\ref{FigRg0Vphi} shows, for
particles identified in a narrow cylindrical shell of radius $R=5\pm
0.5$ kpc at the final time, $t_3$, the azimuthal velocity,
$V_\phi(t_3)$, as a function of the initial guiding center radius,
$R_{\rm g}(t_0)$. This figure shows that stars that have migrated
inward from large radii tend to have larger $V_\phi$ (or,
equivalently, larger $R_{\rm g}$) than those that have migrated
outward from well within the inner disk, leading to a well defined
correlation between rotation velocity and provenance radius. If radius
correlates with metallicity, then a negative metallicity gradient in
the Galaxy would give rise to an anti-correlation between
$V_\phi$ and metal content such as observed.

On the other hand, we note that stars chemically identified to be part
of the Galactic thick disk by their $\alpha$ enhancement \citep[see,
e.g.,][]{Navarro2011} show a gradient in the $V_\phi$ versus [Fe/H]
relation of {\it opposite} sign \citep[see,
e.g.,][]{Lee2011,Kordopatis2013,Recio-Blanco2014}, a result that is
difficult to reconcile in our models assuming a migration-driven
scenario for the origin of the thick disk.  However we notice that in
this scenario the anti-correlation has been reproduced as reported in
Figure 15 of \cite{Loebman2011}. The reason of the differences between
our results and previous results might be in the different models of
galaxies adopted and will be investigated in a forthcoming work.  Note
that the different signs of the gradient suggest a qualitative
distinction in the origin of $\alpha$-poor (thin-disk) stars and
$\alpha$-rich (thick-disk) stars. Such kinematic distinction might
also present an interesting challenge to models that view the Galactic
disk as a single entity made by the overlap of several mono-abundance
populations \citep{Bovy2012a,Bovy2012b}.

Finally, our results suggest an extra consistency check of the migration
interpretation of the thin disk $V_\phi$ versus [Fe/H] anti-correlation: those
that migrate the most should also have lower-than-average vertical velocity
dispersions. In this interpretation, both unusually metal-poor or unusually
metal-rich stars should have lower scaleheights, although the trend might be
more readily noticeable in the metal-poor population (see
Figure~\ref{FigVelDispMig}).  No such effect has been reported in the
literature, but this may be due to difficulties in separating thin-disk from
thick-disk stars at the metal-poor end. Indeed, we note that a kindred effect
has been argued by \citet{Minchev2014} to explain the drop in $\sigma_z$
observed for the most $\alpha$-rich stars in their sample.

\section{Summary and Conclusions}
\label{SecConc}

We have analyzed the effects of a persistent, multi-arm spiral pattern on
the radial distribution of particles in an exponential thin disk. Our disk is
modeled with $10^8$ self-gravitating particles and a rigid halo potential,
thus minimizing the effects of noise. Evolved in isolation, the disk develops
no obvious non-axisymmetric patterns nor is subject to the violent
instabilities that have often plagued $N$-body models of cold stellar disks.

Spiral patterns are seeded in the disk by the temporary addition of massive
perturbers intended to simulate the presence of molecular clouds. The disk
responds to the molecular clouds by developing an intricate multi-arm spiral
pattern that persists , after the perturbers are removed, for over $5$
Gyr. Interestingly, the perturbations cause only minor heating and have no
discernible effect on the vertical structure of the disk. Thus, our
simulations are especially well suited to analyze the effect of long-lived
spiral patterns on the orbits of disk stars.

Stars in the disk respond to the spiral perturbations by migrating
widely through the disk whilst largely preserving the nearly circular
nature of their orbits. Adopting this as a definition of radial
migration, we focus our analysis on the properties of migrators and on
the effects of radial migration on the vertical structure of the
disk. Our main conclusions may be summarized as follows.

\begin{itemize}

\item Radial migration affects a heavily biased subset of disk particles,
  mostly those with small vertical oscillations and close to circular
  orbits. This implies a ``provenance bias'' where migrators sample a very
  select population of stars at their radius of origin. The more a particle
  migrates the stronger the bias, implying that the most extreme migrators
  should be found, at given radius, amongst the stars whose orbits deviate the
  least from the local standard of rest.

\item Migration affects systematically the kinematics of stars; outward
  migrators cool down whereas inward migrators heat up. The magnitude of this
  effect is, however, small compared to the provenance bias referred to above.
  Indeed, although inward migrators heat up as they move in, they come
  from regions that are cooler than their destination and, in addition, they
  constitute a cooler subset of particles from that region. The small amount
  of heating they experience as they move in is not enough to revert the
  bias. In general, migrators have, at given radius, much smaller scaleheights
  and velocity dispersion than non-migrators.

\item Migrators do not ``accommodate'' to match the properties of their new
  surroundings. Rather, they replace particles that were originally resident
  there and that have migrated off. The latter are predominantly stars on very
  nearly circular orbits, and so are the new arrivals.

\item Radial migration might play a substantial role in determining
  the local properties of the thin disk of the Galaxy.  In agreement
  with earlier suggestions, our results suggest that radial migration
  might help to explain the large spread in metallicity observed for
  stars in the ($\alpha$-poor) thin disk as well as the
  anti-correlation observed between metallicity and rotation velocity
  of thin-disk stars.


\end{itemize}

Our results endorse the view that migration might play a substantial
role in the radial redistribution of stars in galaxies that have
experienced sustained spiral structure. Although we have tried to
emphasize the results that seem of general applicability, we also
recognize that our conclusions are based on a single simulation that,
in many ways, fails to capture the fascinating complexity of galaxies
like our own Milky Way. The evolving potential of the forming galaxy;
the accretion of gas and the ongoing formation of stars; the effects
of satellites and substructure; the presence of a bar and of evolving
spirals; all must have left an imprint on the structure of the Galaxy
that our simulations cannot attempt to match.

Further, our simulation adopts a disk that is lighter than that of the Milky
Way and whose spiral patterns, therefore, might differ from that of the Galaxy
\citep[see also][]{Loebman2011}. We cannot therefore rule out that
  other disturbances may give rise to a thick disk, as argued by
  \cite{Roskar2013}. A more massive disk would probably sport spirals with
fewer arms that might lead to faster heating. Although we do not believe that
this would affect qualitatively our conclusions, a more definitive,
quantitative comparison with observation should be attempted with models that
reproduce better the known structural parameters of the Galaxy. Still, we
believe that simplified models like the one we discuss here are useful, since
they allow us to isolate and understand the effects of radial migration so as
to inform the analysis of more ambitious modeling.

\vskip 2em 

This work is funded by NSF grant No. 1211258 and NASA grant
No. 13-ATP13-0053.  ED gratefully acknowledges the support of the
Alfred P. Sloan Foundation. CV-C, ED and JN express their appreciation
towards the Aspen Center for Physics for their hospitality. MA and JN
acknowledge the support of ANPCyT, Argentina by grant No. PICT
2012-1137. This work was supported in part by the National Science
Foundation under grant No. PHYS-1066293. This simulation has been run
on Odyssey at Harvard and on the High Performance Computing cluster
provided by the Advanced Computing Infrastructure (ACI) and Center for
High Throughput Computing (CHTC) at the University of Wisconsin. This
work was supported in part by the National Science Foundation under
Grant No. PHYS-1066293.

\vskip 2em

\bibliographystyle{apj} 
\bibliography{refs}

\end{document}